\documentclass{aa} % for a referee version

\usepackage{natbib}
\usepackage{graphicx}
\begin{document} %
   \title{The fundamental parameters of the roAp star 10~Aql \thanks{Based on observations made with the VEGA/CHARA spectro-interferometer.}}

   % \subtitle{xxx}

\author{K.~Perraut\inst{1}, S. Borgniet\inst{1}, M. Cunha\inst{2}, L. Bigot\inst{3}, I. Brand\~ao\inst{2}, D. Mourard\inst{3}, N. Nardetto\inst{3}, O. Chesneau\inst{3}, H. McAlister\inst{4,5}, T.A.~ten~Brummelaar\inst{5}, J. Sturmann\inst{5}, L. Sturmann\inst{5}, N. Turner\inst{5}, C. Farrington\inst{5}, and P.J. Goldfinger\inst{5}
}

   \institute{Institut d'Astrophysique et de Plan\'etologie de Grenoble, CNRS-UJF UMR 5571,
          414 rue de la Piscine, 38400 St Martin d'H\`eres, France
          \and
             Centro de Astrof\'\i sica e Faculdade de Ci\^encias, Universidade do Porto, Portugal
             \and
             Laboratoire Lagrange, UMR 7293 UNS-CNRS-OCA, Boulevard de l'Observatoire, BP 4229, 06304 Nice cedex 4, France
          \and Georgia State University, P.O. Box 3969, Atlanta GA
         30302-3969, USA
         \and CHARA Array, Mount Wilson Observatory, 91023 Mount Wilson CA, USA
         }

         \offprints{} \date{Received
         ...; accepted ...}

  \abstract
   {Owing to the strong magnetic field and related abnormal surface layers existing in rapidly oscillating Ap (roAp) stars, systematic errors are likely to be present when determining their effective temperatures, which potentially compromises asteroseismic studies of this class of pulsators.}
   {Using the unique angular resolution provided by long-baseline visible interferometry, our goal is to determine accurate angular diameters of a number of roAp targets, so as to derive unbiased effective temperatures ($T_{\rm eff}$) and provide a $T_{\rm eff}$ calibration for these stars.}
   {We obtained long-baseline interferometric observations of 10~Aql with the visible spectrograph VEGA at the combined focus of the CHARA array. We
    derived the limb-darkened diameter of this roAp star from our visibility measurements. Based on photometric and spectroscopic data available in the literature, we estimated the star's bolometric flux and used it, in combination with its parallax and angular diameter, to determine the star's luminosity and effective temperature.}
   {We determined a limb-darkened angular diameter of $0.275 \pm 0.009$~mas and deduced a linear radius of $R$~=~2.32~$\pm$~0.09~${\rm R_{\odot}}$. For the bolometric flux we considered two data sets, leading to an effective temperature of $T_{\rm eff} = 7800 \pm 170$~K and a luminosity of $L/{\rm L}_\odot = 18 \pm 1$ or $T_{\rm eff} = 8000 \pm 210$~K and $L/{\rm L}_\odot = 19 \pm 2$. Finally we used these fundamental parameters together with the large frequency separation determined by asteroseismic observations to constrain the mass and the age of 10~Aql, using the CESAM stellar evolution code. Assuming a solar chemical composition and ignoring all kinds of diffusion and settling of elements, we obtained a mass M/${\rm M_{\odot}} \sim$~1.92  and an age of $\sim$~780~Gy or a mass M/${\rm M_{\odot}} \sim$~1.95 and an age of $\sim$~740~Gy, depending on the derived value of our computing the bolometric flux.}
   {For the first time, thanks to the unique capabilities of VEGA, we managed to determine an accurate angular diameter for a star smaller than 0.3~mas and to derive its fundamental parameters. In particular, by only combining our interferometric data and the bolometric flux, we derived an effective temperature that can be compared to those derived from atmosphere models. Such fundamental parameters can help for testing the mechanism responsible for the excitation of the oscillations observed in the magnetic pulsating stars.}


    \authorrunning {K. Perraut et al.}
   \titlerunning{}
   \maketitle
%________________________________________________________________

\section{Introduction}

Rapidly oscillating Ap (roAp) stars are a subgroup of Ap stars that are chemically peculiar main-sequence magnetic stars of spectral type B8 to F0.
The roAp stars exhibit strong large-scale organized magnetic fields (between a hundred of G up to 24.5~kG for HD~154708 -- \cite{HD154708}), abundance
anomalies of { Sr, Cr and} rare earths, small rotation speeds ($\leq$~100~km/s and often $\leq$~30-40~km/s), and pulsations with a range of amplitudes
from about 10~millimagnitudes \citep{kurtzetal06} to 10~micromagnitudes \citep{balonaetal11a} and periods ranging from 6~min for HD~134214 \citep{Saio2012} up to 24~min for HD~177765 \citep{Alentiev}. These pulsations are interpreted as high-order, low-degree acoustic modes.

%The "Ap phenomenon" is closely related to the strong magnetic field and may be related to a certain phase of the star life \citep{HubrigroApnoAp}.
The mechanism inducing the slow rotation of these stars is likely due to magnetic braking. \citet{Stepien} concludes that this magnetic braking must occur during the pre-main sequence phase so as to reproduce the angular momentum observations of the Ap stars. This theory is corroborated by the recent study of the rotation of magnetic Herbig AeBe stars by \citet{Alecian2012}. In addition, several spectro-polarimetric observations of Herbig AeBe provide evidence of magnetic
fields whose strength and topology are similar to those of Ap stars (\cite{Wade}; \cite{Alecian08}). This seems to indicate that magnetic Herbig AeBe stars are the progenitors of Ap stars.

In (ro)Ap stars, the strong magnetic field is supposed to play a key role in the abundance distribution by stabilizing the external layers and partially
or fully suppressing the convection. Vertical and horizontal microscopic diffusion of the chemical elements \citep{Michaud} is thus believed to occur, leading to spotted surfaces that have been reconstructed by Doppler or Zeeman-Doppler imaging for several (ro)Ap stars (\cite{Luft2003}; \cite{Koch2004}; \cite{HD24712}; \cite{Koch2010}). Such abnormal surface layers might generate systematic errors when determining stellar luminosities and effective temperatures by spectrometric or photometric techniques \citep{Matthews}.

For a long time it was considered that such a strong magnetic field prevented oscillations from existing. But in 1978, 12-minute oscillations were first detected in the Ap Przybylski's star \citep{Kurtz78} which became the first roAp star. Up to now, more than 40 roAp have been disco\-vered. Their oscillations are different from those observed in the other types of pulsating stars that lie in the instability strip of the Hertzsprung-Russell (HR) diagram (like $\delta$-Scuti, $\gamma$-Dor, and Cepheids stars). While the periods of the roAp oscillations are analogous to those of the solar ones (supposed to come from the Lighthill process), their amplitudes cover a much wider range. The mechanism res\-ponsible for the excitation of the oscillations observed
in roAp stars is still a matter of debate. Currently the most promising theory is that envelope convection is suppressed by the strong magnetic fields present in these stars and excitation  then occurs as a result of the opacity mechanism acting on the convectively stable hydrogen ionization zone \citep{balmforthetal01}. This hypothesis has been tested with reasonable success through nonadiabatic calculations \citep{Cunha2002}. Moreover, microscopic diffusion
is also supposed to have a significant impact on the excitation process \citep{Theado}.

Deriving accurate fundamental parameters of a large sample of roAp stars is of strong interest for comparing both adiabatic and nonadiabatic model predictions with observations. \citet{Creevey} and \citet{2007A&ARv..14..217C} have shown that an accurate linear radius provided by interferometry combined with asteroseismic data allows an accurate stellar mass to be derived and stellar interiors to be tested: typically an accuracy better than 3\% on the radius allows an accuracy better than 4\% to be reached on the mass. With the recent operation of visible interfe\-rometers working on very long-baseline arrays like VEGA on the CHARA array \citep{2009A&A...508.1073M}, the angular resolution is now of the order of 0.2 millisecond of arc (mas), which allows resolving several roAp stars despite their very small angular size (i.e. $\leq$~1~mas). Up to now, the angular diameters of three roAp stars have been determined by optical interfero\-metry: a limb-darkened angular diameter of 1.105~$\pm$~0.037~mas for $\alpha$~Cir with SUSI \citep{Bruntt2008}, limb-darkened angular diameters of 0.669~$\pm$~0.017~mas and 0.415~$\pm$~0.017~mas for the two components of $\beta$~CrB with FLUOR \citep{Bruntt2010}, and a limb-darkened angular diameter of 0.564~$\pm$~0.017~mas for $\gamma$~Equ with VEGA \citep{2011A&A...526A..89P}. These angular diameters have been used to derive the stars' effective temperatures and differences between these \textit{interferome\-tric} effective tempe\-ratures and the \textit{photometric} effective temperatures have been highlighted.

Thanks to the unique capabilities of VEGA in terms of high angular resolution, we can extend this sample to smaller roAp stars. We present here the results we obtained on 10~Aql (HD~176232; F0p), which is one of the brightest roAp stars of the northern hemisphere ($m_V$~$\sim$~5.9). Its parallax is well known, $\pi_P$~=~12.76~$\pm$~0.29~mas \citep{Hipparcos}. Its rotation period is unknown but it might be of several hundreds of years, given its rotation speed $v \sin i$ less than 2.0~$\pm$~0.5~km/s (\cite{Koch2002}). \citet{ryabchikova00} and, more recently, \citet{Nesvacil} have studied the abundances of this target. \citet{ryabchikova00} detect strong overabundances in doubly ionized elements of rare earths (like Neodymium or Praseodymium), while \citet{Nesvacil} derive a self-consistent, chemically stratified atmosphere model for 10~Aql. Elements Mg and Co were found to be the least stratified, while Ca and Sr exhibited the largest abundance stratifications. \citet{Koch2002} have measured a mean magnetic field modulus of $\langle B \rangle$~=~1.5~$\pm$~0.1~kG in 10~Aql from Zeeman splitting.

This star was discovered as a roAp by \cite{Heller} with a pulsation period of about 11.9 min and an amplitude well below one millimagnitude. \cite{Huber} continuously followed this target with the MOST satellite during one month for identifying the pulsation frequencies and studying the effect of the magnetic field. Even if the frequency spectrum was found to not be rich enough to allow the authors to converge towards a unique model, they could derive a large frequency separation of $\Delta \nu$~=~50.95~$\mu$Hz and show that the best fit corresponds to a 1.95~${\rm M_{\odot}}$ model ha\-ving solar metallicity, suppressed envelope convection, and homogenous helium abundance.

In this paper, we report our interferometric observations of 10~Aql with the VEGA instrument (Sect.~\ref{sec:obs}), which allow a limb-darkened angular diameter to be derived. To fix the star's position in the HR diagram, we compute the star's bolometric flux (Sect.~\ref{flux}) and derive its fundamental parameters (Sect.~\ref{fund}). We then use a stellar evolution code to determine the mass and the age of the target (Sect.~\ref{CESAM}). Finally, we discuss how our results compare with those pu\-blished in the literature (Sect.~\ref{discussion}).

\section{Interferometric observations and data processing}
\label{sec:obs}

\subsection{CHARA/VEGA observations}

\begin{table*}[t] \centering \caption{Log of the observations. Columns 1, 2, and 3 give the date, the UT time, and the hour angle (HA) of the observations. Columns 4, 5, and 6 give the telescope pair, the projected baseline length $B_{\rm p}$ and its orientation $PA$. Column 7
gives the central wavelength of the spectral range used for computing the squared visibility. Columns 8, 9, and 10 give the calibrated
squared visibility $V^2$, the statistic error on $V^2$, and the systematic error on $V^2$ (see text for details). The last column provides
the Fried parameter for each observation.} \label{tab:log}
\label{tab:log}
\begin{tabular}{ccccccccccc}
\hline Date & UT & HA & \multicolumn{3}{c}{Baselines} & $\lambda_0$ & $V^2$ & $\sigma_{\rm stat}$ & $\sigma_{\rm syst}$ & $r_0$ \\
& (h) & (h) & Telescopes & $B_{\rm p}$ (m) & $PA$ ($^\circ$) & (nm) & & & & (cm)\\
\hline
24-07-2011 & 6.5 & -0.2 & E1E2 & 65 & -120 & 715 & 0.98 & 0.07 & 0.004 & $\sim$~7\\
%           &     &     & E2W2 & 155 & -114 & 715 & 0.80 & 0.069 & 0.019 & KO\\
           &     &     & E2W2 & 155 & -114 & 715 &  &  &  & \\
           &     &     & E1W2 & 219 & -116 & 715 & 0.58 & 0.06 & 0.03 &\\
24-07-2011 & 6.5 & -0.2 & E1E2 & 65 & -120 & 695 & 0.97 & 0.09 & 0.004 & $\sim$~7\\
%           &     &     & E2W2 & 155 & -114 & 695 & 0.73 & 0.068 & 0.018 & KO\\
           &     &     & E2W2 & 155 & -114 & 695 &  &  &  & \\
           &     &     & E1W2 & 219 & -116 & 695 & 0.66 & 0.08 & 0.03 &\\
\hline
01-09-2011 & 4.4 & 0.2 & E1E2 & 63 & -122 & 735 & 0.86 & 0.05 & 0.003 & 6-7 \\
%           &     &     & E2W2 & 152 & -116 & 735 & 0.69 & 0.057 & 0.014 & KO\\
           &     &     & E2W2 & 152 & -116 & 735 &  &  &  & \\
%           &     &     & E1W2 & 215 & -118 & 735 & 0.50 & 0.047 & 0.022 & KO\\
           &     &     & E1W2 & 215 & -118 & 735 &  &  &  & \\
           &     &     & E1E2 & 63 & -122 & 715 & 0.96 & 0.06 & 0.004 &6-7 \\
%           &     &     & E2W2 & 152 & -116 & 715 & 0.74 & 0.062 & 0.016 & KO\\
           &     &     & E2W2 & 152 & -116 & 715 &  &  &  & \\
%           &     &     & E1W2 & 215 & -118 & 715 & 0.47 & 0.049 & 0.022 & KO\\
           &     &     & E1W2 & 215 & -118 & 715 & & & & \\
\hline
%21-05-2012 & 10.1 & -0.8 & W1W2 & 101 &  100 & 730 & 1.11 & 0.053 & 0.011 & 7-10 KO\\
21-05-2012 & 10.1 & -0.8 & W1W2 & 101 &  100 & 730 & & & & 7-10\\
           &      &      & E1W2 & 222 & -113 & 730 & 0.65 & 0.04 & 0.03 & \\
           &      &      & E1W1 & 311 & -103 & 730 & 0.46 & 0.02 & 0.04 & \\
%           &      &      & W1W2 & 101 &  100 & 710 & 1.26 & 0.047 & 0.013 & KO\\
           &      &      & W1W2 & 101 &  100 & 710 &  & &  & \\
%           &      &      & E1W2 & 222 & -113 & 710 & 0.71 & 0.080 & 0.034 & KO\\
           &      &      & E1W2 & 222 & -113 & 710 &  &  &  & \\
           &      &      & E1W1 & 311 & -103 & 710 & 0.43 & 0.03 & 0.04 & \\
\hline
16-07-2012 & 5.1 & -2.16 & W1W2 & 83 & 107 & 730 & 0.95 & 0.07 & 0.007 & $\sim$~10 \\
           &     &         & E1W2 & 211 & 71 & 730 & 0.80 & 0.06 & 0.03 & \\
           &     &         & E1W1 & 282 & 81 & 730 & 0.56 & 0.07 & 0.05 & \\
16-07-2012 & 5.9 & -1.35 & W1W2 & 95 & 102 & 730 & 0.87 & 0.05 & 0.008 & $\sim$~8 \\
           &      &        & E1W2 & 220 & 68 & 730 & 0.70 & 0.06 & 0.03 & \\
           &      &       & E1W1 & 304 & 79 & 730 & 0.55 & 0.08 & 0.05 & \\
\hline
\end{tabular}
\end{table*}

The CHARA array \citep{chara} hosts six one-meter telescopes arranged in
a Y shape and oriented to the east (E1 and E2), south (S1 and S2) and west
(W1 and W2). The baselines $B$ span between
30~m (S1S2) and 330~m (S1W1) allowing a maximal angular resolution of $\lambda/(2 B)\sim$~0.2~mas
to be reached in the visible range. The VEGA spectrograph is
one of the visible focal instruments of the array. It can combine 2, 3, or 4
telescopes, and it allows recording spectrally dispersed
fringes from 0.45~$\mu$m to 0.85~$\mu$m at a spectral resolution of 6000 (medium
resolution) or 30000 (high resolution) as described in \cite{2009A&A...508.1073M}.
VEGA is equipped with two photon-counting detectors looking at two different spectral
bands. In the medium spectral resolution, these two bands have spectral bandwiths of
30~nm for the shortest observed wavelengths and 40~nm for the longest ones. The two bands are sepa\-rated by about 170~nm. The medium spectral resolution is used for diameter determination via squared visibility measurements. This mode
reaches a limiting {correlated magnitude} of $m_{\rm V} \sim$~6.5 in medium seeing conditions. The high spectral resolution is used for
kinematics studies via differential observable measurements. The limiting {correlated magnitude} is then $m_{\rm V} \sim$~4.5.

VEGA can be operated in parallel with the CLIMB beam combiner working in the K band and acting as a coherence sensor \citep{CLIMB}. In this near-infrared range, the turbulence is less penalizing and the fringe contrast is higher since the angular resolution is smaller. CLIMB can thus ensure the necessary fringe stabili\-ty required for the long observing sequences in the visible. We measured a typical residual jitter on the optical path difference on the order of 7~$\mu$m, which is in good agreement with our need in medium spectral resolution (20~$\mu$m).

\begin{figure*}[t]
  \begin{center}
  \includegraphics[width=9.7cm, angle=0]{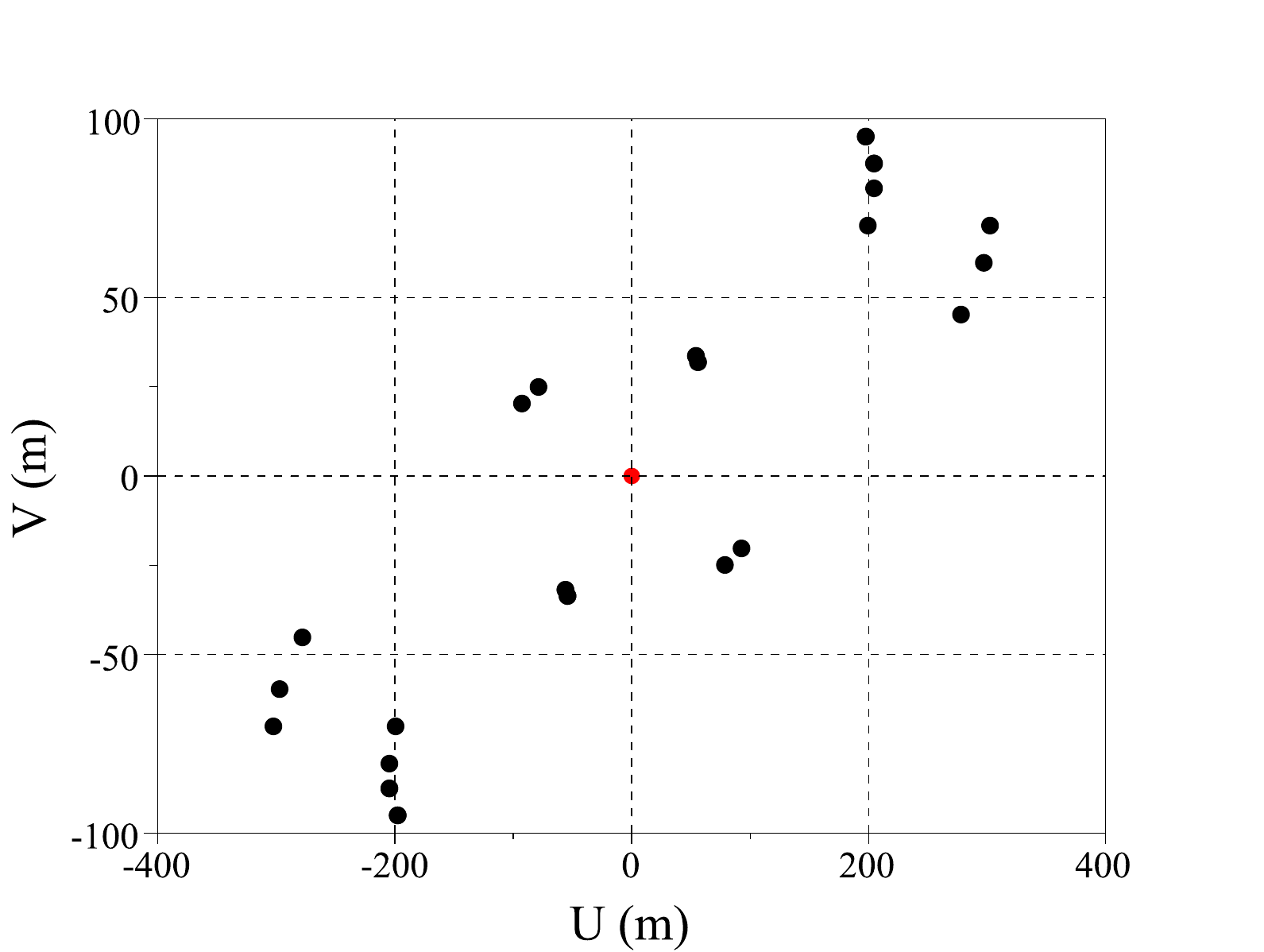}~\includegraphics[width=9.7cm, angle=0]{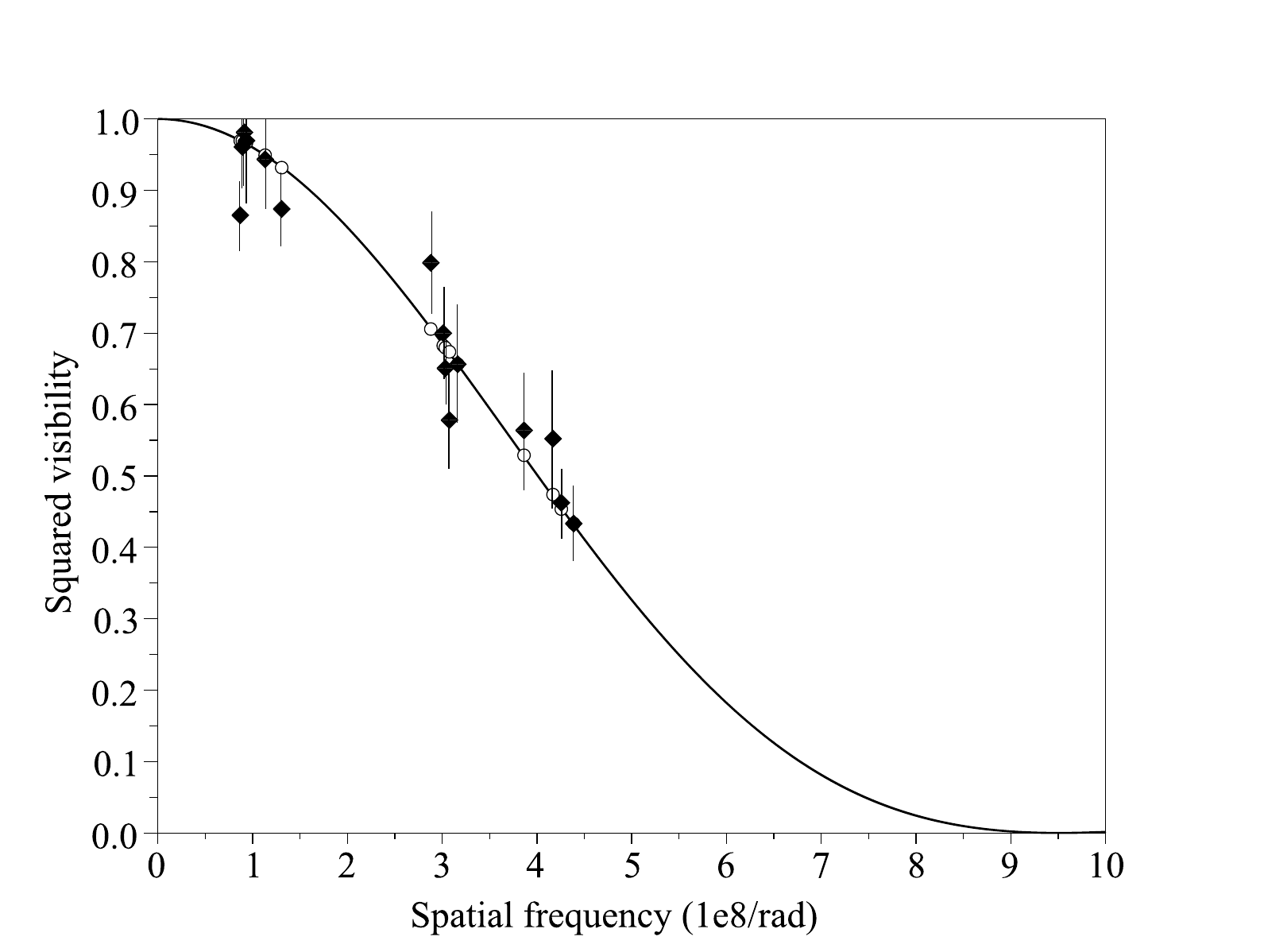}
  \end{center}
  \caption{\textit{Left.} Spatial frequency coverage of the VEGA observations. \textit{Right.} Squared visibility versus spatial frequency for 10~Aql obtained with the VEGA observations (diamonds). The solid line
 and the open circles represent the uniform-disk best model provided by LITPRO.}
\label{fig:V2}
\end{figure*}

We have observed 10~Aql with different telescope triplets chosen to partially resolve the small expected angular diameter of the target (about 0.3~mas, so close to the angular resolution limit of VEGA). The spectral bands were centered on 550~nm and 720~nm. For the shorter wavelengths, only the excellent seeing conditions allow the visibility to be retrieved. We followed a sequence calibrator-target-calibrator, with 40 or 60 blocks of 1000 short exposures (of 25 ms) per star. Swapping from one star to ano\-ther every 20 or 30 minutes ensures that the instrumental transfer function is stable enough. We used the SearchCal software \citep{searchcal} proposed by the JMMC\footnote{Available at www.jmmc.fr/searchcal} to find two relevant calibrators for our target: HD~170878 and HD~160765. Their uniform-disk angular diameters are determined by surface-brightness versus color-index relationships. An accuracy of 7\% on the angular diameter is obtained by using the (V, V~-~K)  polynomial relation. This leads to a uniform-disk angular diameter in the R band of 0.242~$\pm$~0.017~mas for HD~170878 and of 0.151~$\pm$~0.011~mas for HD~160765. The observation log is given in Table~\ref{tab:log}, and the corresponding spatial frequency (u, v) coverage is displayed to the left in Figure~\ref{fig:V2}.

\subsection{VEGA data processing}
\label{diameter}

The interferometric data were reduced using the standard VEGA reduction pipeline described in \citet{2009A&A...508.1073M}. This software allows computing the squared visibility in wide spectral bands based on the spectral density analysis. For each night, we first checked that the instrumental transfer function is stable over the night by computing this function transfer for all the calibrators of all the programs of the night as explained in \citet{2012SPIE-VEGA}. Bad sequences due to poor seeing conditions or instru\-mental instabilities noted in the journal of the night were rejected.

\begin{figure}[h]
\begin{center}
\includegraphics[angle=0,width=9cm]{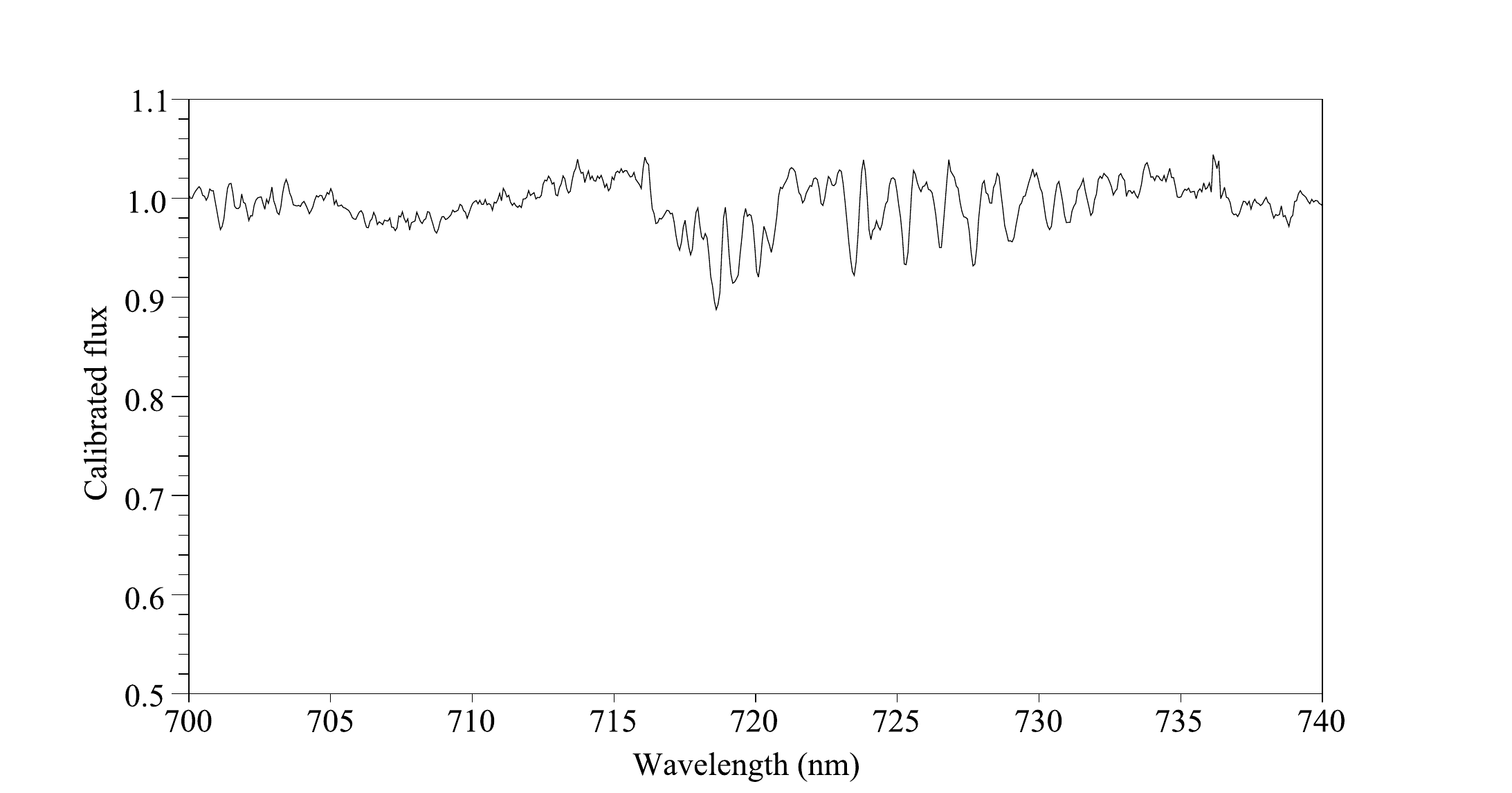}
\end{center}
\caption{Spectrum of 10~Aql at the medium spectral resolution of VEGA ($\sim$~6000).
Each visibility measurement is computed over a 20-nm range.}
\label{fig:spectralband}
\end{figure}

For each data sequence (calibrator-target-calibrator), we computed the raw squared visibility for each block of 1000 individual frames and for different spectral bands of 20~nm (whose central wavelengths $\lambda_0$ are given in Table~\ref{tab:log}). Such large bandwidths lead to an effective spectral resolution of 36 (at the optical wavelength of VEGA), which means that our visibility measurements are mainly sensitive to the continuum (or photosphere) of the star. The individual spectral lines of 10~Aql are indeed unresolved at such a spectral resolution (Fig.~\ref{fig:spectralband}).

Using the known angular diameter of the calibrators, we calibrated the target squared visibilities $V^2$ and estimated a weighted-mean calibrated squared visibility and the corresponding errors that were twofold: $\sigma_{\rm stat}$ was derived from the statistical dispersion over the block's measurements, and $\sigma_{\rm syst}$ accounted for the uncertainty on the calibrator diame\-ter. For each target sequence, we then studied the influen\-ce of selecting the individual block's squared visibi\-lity on a signal to noise (S/N) criterion. We determined the threshold on S/N where the estimated calibrated visibility started to be biased. Under good seeing conditions, putting a high S/N threshold does not change the mean value of the squared visibilities and only decreases the statistical dispersion $\sigma_{\rm stat}$, as for the data of May 2012. However, for ave\-rage or bad seeing conditions, putting a high S/N thre\-shold tends to increase the squared visibilities and bias the measurements. Such a trend clearly appeared for the data recorded in September 2011 with a $r_0$ of 6-7~cm. For such seeing conditions we could only retrieve unbiased squared visibilities for the shortest baseline (i.e., for the most contrasted fringes).

From this study of our measurement bias, we decided to consider furthermore the values of the calibrated squared visibilities obtained for a S/N threshold of four and discarded the biased visibilities. We kept 15 visibi\-lity points whose values are given in Table~\ref{tab:log}. For baselines up to about 200~m the error on the squared visibilities is dominated by the statistical dispersion ($\sigma_{\rm stat}$ is 3 or 4 times larger than $\sigma_{\rm syst}$), while for the longest baseline (i.e., more than 300~m), it can be dominated by the accuracy on the calibrators' diameter. This is the case for the excellent data of May 2012 where $\sigma_{\rm syst}$ is twice greater than $\sigma_{\rm stat}$ for the E1W1 baseline.

\subsection{Deriving a uniform-disk angular diameter}

We plotted all the calibrated squared visibilities as a function of the spatial frequency ($B_p/\lambda$) and performed a model fitting using LITpro \citep{LITPRO} proposed by the JMMC\footnote{$\textrm{www.jmmc.fr/litpro\_page.htm}$}.
This fitting engine is based on a modified Levenberg-Marquardt algorithm combined with the trust region method. The software provides a user-expandable set of geometrical elementary
models of the object, combinable as building blocks. The fit of all the visibility measurements versus spatial frequency leads to a uniform-disk angular diameter $\theta_{\rm UD}$ of 0.264~$\pm$~0.0085~mas for 10~Aql (Fig.~\ref{fig:V2}-right). We now check the effect of the wavelength by independently fitting the visibility measurements corresponding to 730-735 nm (8 visibility points) and to 695-715 nm (7 visibility points). The angular diameter we obtain at 730-735~nm is slightly larger than the one at 695-715~nm, but the difference (less than 1-$\sigma$) is not significant compared to the error bars. Moreover, both values are also consistent with the result from the global fit. On the other hand, owing to the (u, v) coverage (Fig.~\ref{fig:V2}-left)), we only probe one baseline direction (around -105$^\circ$~$\pm$~15$^\circ$), and we cannot determine any angular diameter variation with respect to the baseline orientation. As a consequence, we consider all measurements as a whole and fix $\theta_{\rm UD}$ of 0.264~$\pm$~0.0085~mas in the following.

\subsection{Deriving a limb-darkened angular diameter}

The tables of \cite{TableClaret} provide the linear limb-darkening coefficients $u(\lambda)$
in the R, I, J, H, and K bands used to determine the limb-darkened angular diameter in the corresponding band.
We derived the limb-darkened angular diameter $\theta_{\rm LD}$ of 10~Aql in the R band through the formula
\begin{equation}\label{eq1}
    \theta_{\rm LD} = \theta_{\rm UD} \sqrt{\frac{1 - \frac{u(R)}{3}}{1 - 7 \frac{u(R)}{15}}}
\end{equation}
where $u(R)$ denotes the limb-darkening coefficient in the R band. It is obtained by the adjustment of the radial intensity distribution on the stellar disk using a stellar atmosphere model characterized by the effective temperature
($T_{\rm eff}$), the gravity ($\log g$), and the metallicity ([Fe/H]).
\noindent For a given $\log g$, we computed the coefficients $u(R)$ and the corresponding limb-darkened diameters when the effective
temperature spans from 7250~K to 8250~K. We tested three different $\log g$: 3.5, 4, and 4.5. Over these huge temperature and gravity ranges, the limb-darkened diameter only varies by 1.2~$\times$~10$^{-3}$~mas for $\log g$ equal to 4 and 4.5 and of
1.7~$\times$~10$^{-3}$~mas for $\log g$~=~3.5. {We also tested different metallicities from -5 to +1: over this range the limb-darkened diameter only varies of 10$^{-3}$~mas.} These variations are 5-7 times smaller than our error bar on the uniform-disk
diameter. We fixed $u(R)$~=~0.505 (for $T_{\rm eff}$~=~7750~K, $\log g$~=~4{, and a solar metallicity}) and obtained a limb-darkened angular
diameter of $\theta_{\rm LD}$~=~0.275~$\pm$~0.009~mas (i.e., about 3\% of relative accuracy). To our knowledge, this corresponds
to the smallest angular diameter ever measured.

\section{Computation of the bolometric flux}
\label{flux}

To determine the fundamental parameters of 10~Aql, in particular its effective temperature and luminosity, we need to compute the star's bolometric flux, in addition to the stellar angular diameter derived in Sect.~\ref{diameter}. This, in turn, requires knowledge about the star's spectral energy distribution.

\subsection{Data}

The apparent flux distribution for 10~Aql was obtained by combining photometric and spectroscopic data available in the literature. Since a significant amount of data is availa\-ble, part of which overlapping in wavelength, we considered two different data combinations for deriving the bolometric fluxes, which provided an indication of uncertainties not accounted for in the individual data measurements.

For the first flux distribution, hereafter {\it case~1}, we considered the following data. \begin{itemize}
\item For wavelengths in the range 1150~\AA~$< \lambda <$~3349~\AA ,  we used two rebinned spectra from the Sky Survey Telescope obtained at the \textit{IUE} ``Newly Extracted Spectra'' (INES)
data archive\footnote{http://sdc.laeff.inta.es/cgi-ines/IUEdbsMY}.
Based on the quality flag listed in the IUE spectra \citep{gar97},
we removed all bad pixels from the data, and we also removed
the measurements with negative flux.
\item For wavelengths in the range 3349~\AA~$< \lambda <$~10198~\AA, we used a low-dispersion STIS spectra extracted from the Next Generation Spectral Library (NGSL), 2nd version\footnote{http://archive.stsci.edu/prepds/stisngsl}.
\item At longer wavelengths, we collected the photometric data from the 2MASS All Sky Catalog of point sources\footnote{http://irsa.ipac.caltech.edu/applications/Gator/} \citep{cutri03}. Filter responses and zeropoints were taken from \cite{cohen03}.

\end{itemize}

\begin{figure*}[t]
\centering
\includegraphics[width=9cm, angle=0]{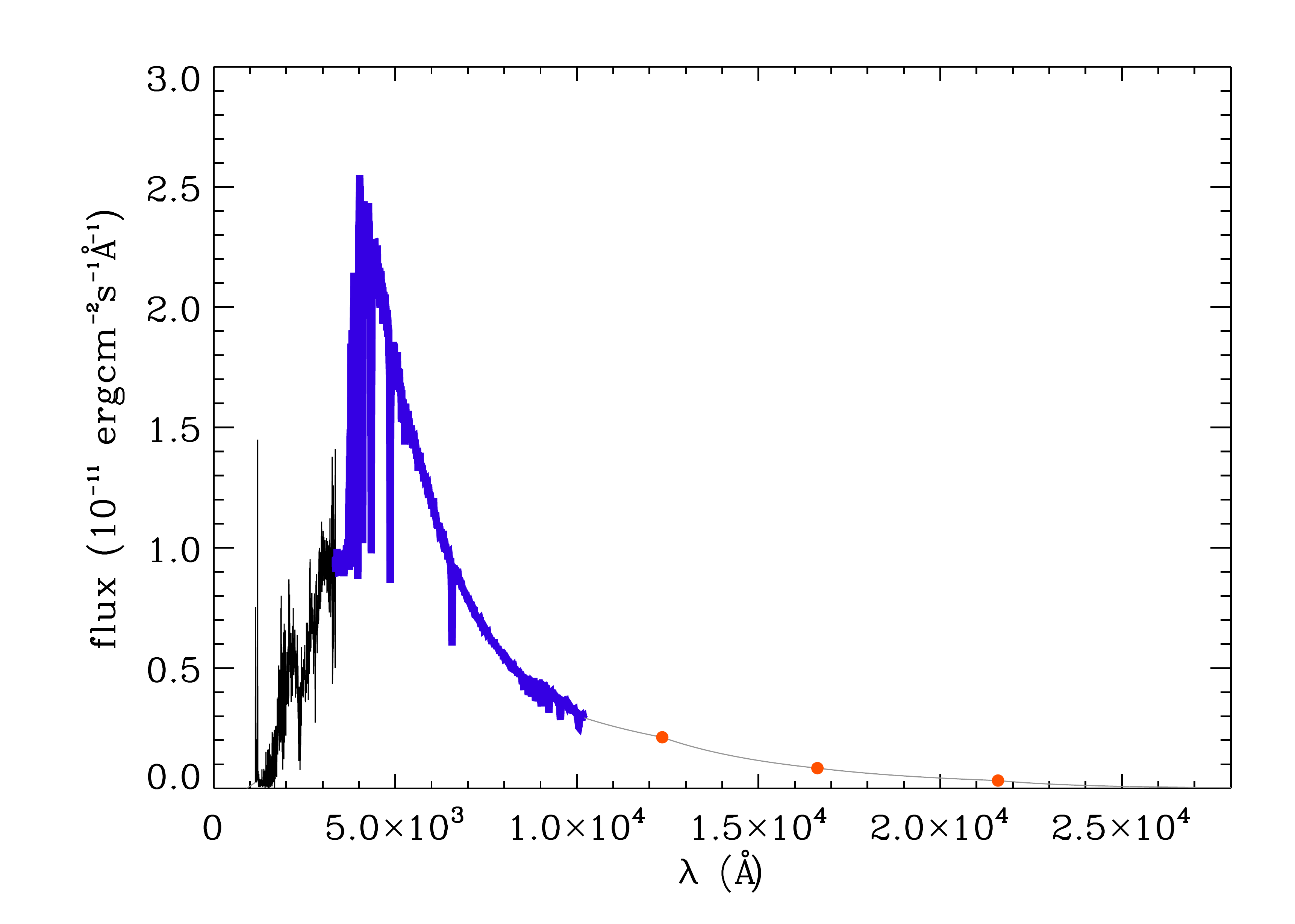} \includegraphics[width=9cm, angle=0]{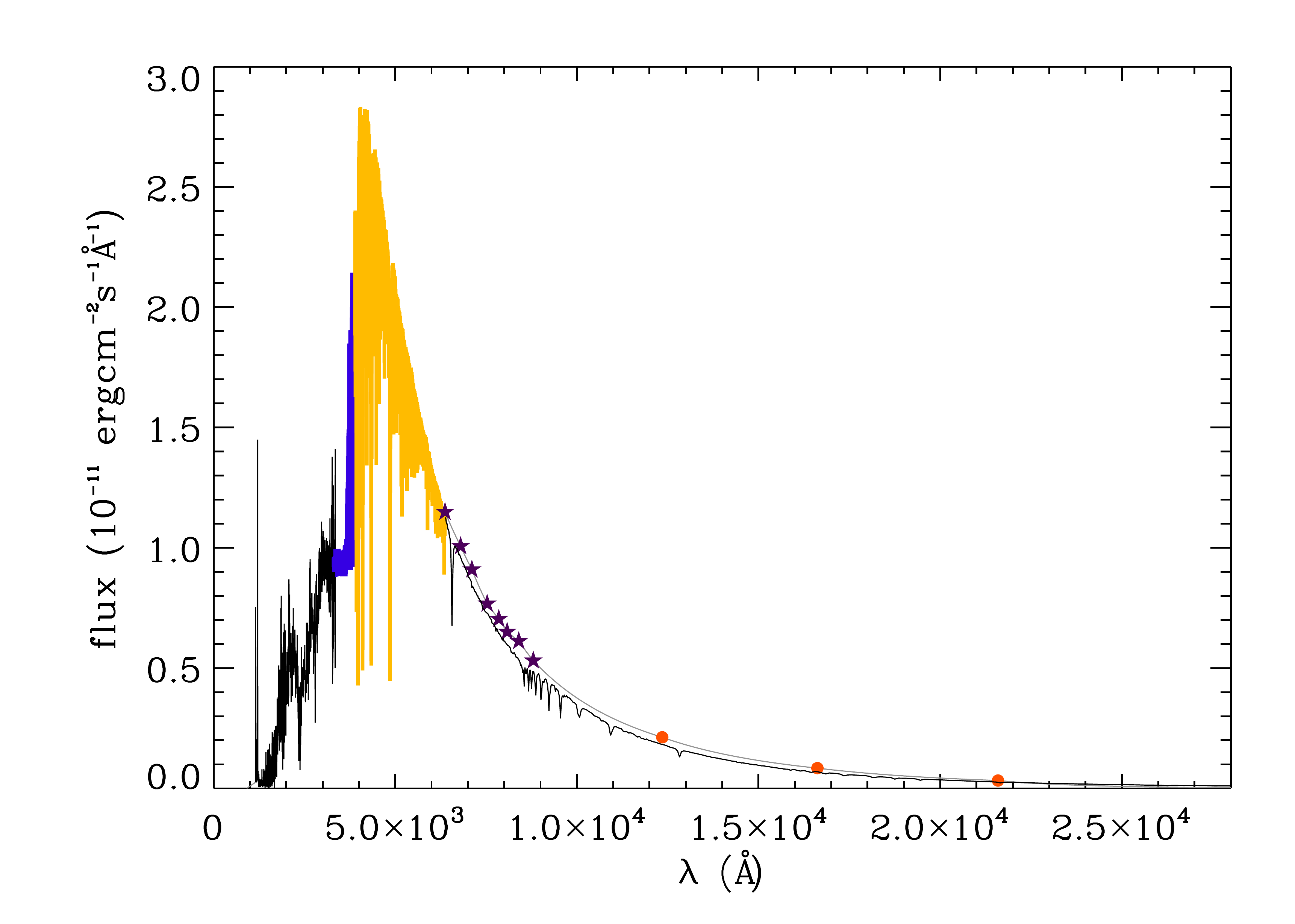}
\caption{Spectral energy distribution obtained for 10Aql using the dataset described in  {\it case~1} (left panel) and the dataset described in   {\it case~2} (right panel). In both plots, the IUE spectrum is shown in black at the lower wavelengths, the STIS spectrum is shown in blue, the interpolations are shown in gray and the 2MASS data are shown by orange circles. In addition, in the right panel, the ELODIE spectrum is shown in yellow, the Kurucz model at the higher wavelengths is shown in black and the Breger data are shown by purple stars. [See online edition for a color version.]}
  \label{SED}
\end{figure*}

For the second flux distribution, hereafter {\it case~2}, we considered the same data as above for the shorter wavelengths in the range  1150~\AA~$< \lambda <$ 3900~\AA, as well as for the longer wavelengths covered by the 2MASS photometry. For wavelengths in between, the data above were replaced by the following data: \begin{itemize}
\item For wavelengths in the range 3900~\AA~$< \lambda <$~6370~\AA,  we used a flux-calibrated ELODIE spectra extracted from the ELODIE.3.1 library\footnote{www.obs.u-bordeaux1.fr/m2a/soubiran/elodie\underline{\hspace{0.1cm}}library.html} \citep{prugniel01,prugniel07}.
\item For wavelengths between 6370~\AA~and 8800~\AA, we collected spectrophotometric data from \cite{breger76}.
\end{itemize}

In both cases we performed linear interpolations (on logarithmic scale) between the photometric data points, as well as at both extremes of the spectral distribution, namely between 912\,\AA~and 1150\,\AA, considering zero flux at 912\,\AA, and between 21590\,\AA~and 1.6~$\times 10^6$\,\AA, considering zero flux at 1.6~$\times 10^6$\,\AA.

\subsection{Determination of $f_{\rm bol}$}

The bolometric flux, $f_{\rm bol}$, was then computed from the integral of the spectral energy distribution found for each case. The spectral energy distributions are shown in Fig.~\ref{SED}, and the corresponding bolometric fluxes are presented in Table~\ref{tab:res}, as are the uncertainties in the two values of the bolometric flux. These uncertainties were estimated by considering the uncertainties quoted in the data sources (when observed calibrated spectra are used), and a conservative uncertainty of 15$\%$ (when interpolations or extrapolations are made). In particular, the uncertainties quoted in the refe\-rences cited above led us to take $10\%$ uncertainty on the flux component computed from the combined IUE spectra, 3$\%$ uncertainty on the flux component computed from the STIS spectrum, and 2.5$\%$ uncertainty on the flux component computed from the ELODIE spectrum.

The contributions from the extremes of the spectral distribution (corresponding to extrapolations) were found to account for less than 1$\%$ of the total flux. Moreover, we have checked that the use of interpolation between photometric data is adequate by comparing, for {\it case~2}, the flux obtained through the approach described above with the flux obtained when using, for wavelengths larger than 6370~\AA, a Kurucz model computed with ATLAS9 \citep{castelli04} chosen to fit the star's spectrum in the visi\-ble and the stars' photometry in the infrared (see pre\-vious works, e.g., for details \cite{huber2012}). The relative difference in the fluxes obtained through these two procedures for {\it case~2} was found to be smaller than 3$\%$, so, well within the computed uncertainties.

\begin{table*}[t]
\caption{Fundamental parameters computed for 10Aql from two data sets and assuming a solar chemical composition (see text for details).}
\centering
\begin{tabular}{cccccc}
\hline
Datasets & $f_{\rm bol}$ (erg~cm$^{-2}$~s$^{-1}$) & $T_{\rm eff}$ (K) &$ L$ (${\rm L}_\odot$) &$ M$ (${\rm M}_\odot$) & Age (Gy)\\
\hline
$case~{\it 1}$ & $[0.93\pm 0.05]\times$10$^{-7}$ & 7800 $\pm$ 170 & 18 $\pm$ 1 & ${\rm 1.92 \pm 0.03}$ &  $780 \pm 40 $\\
\hline
$case~{\it 2}$ & $[1.02 \pm 0.08]\times$ 10$^{-7}$ & 8000 $\pm$ 210 & 19 $\pm$ 2 & ${\rm 1.95 \pm 0.05}$ &  $740 \pm 60$ \\
\hline
\end{tabular}
\label{tab:res}
\end{table*}

\section{Determination of the fundamental parameters}
\label{fund}

\subsection{Linear radius}

From a simple Monte Carlo simulation, we derived the radius of 10~Aql and its error thanks to the formula
\begin{equation}
    R \pm \delta R = \frac{\theta_{\rm LD} + \delta \theta_{\rm LD}}{9.305 \times (\pi_{\rm P} + \delta \pi_{\rm P})},
\end{equation}
where $R$ stands for the stellar radius (in solar radius, ${\rm R_{\odot}}$), $\theta_{\rm LD}$ for the limb-darkened angular diameter (in mas),
and $\pi_{\rm P}$ for the parallax (in second of arc).
We obtained $R$~=~2.32~$\pm$~0.09~${\rm R_{\odot}}$. Our accuracy determination of 3.3\% on the angular diameter coupled with the well-known paral\-lax (at a 2.3\%-precision level) allows us to reach an accuracy of 3.9\% on the radius determination.

\subsection{Effective temperature and luminosity}

We used the measured angular diameter and bolometric flux to estimate the effective temperature
from the definition,
\begin{equation}\label{eq1}
\sigma T_{\rm eff}^{4} = 4 f_{\rm bol}/\theta_{LD}^{2},
\label{eqteff}
\end{equation}
where $\sigma$ stands for the Stefan-Boltzmann constant ($5.67\times\,10^{-5}$~erg~cm$^{-2}$~s$^{-1}$~K$^{-4}$).

The star's luminosity was computed by combining the parallax and bolometric flux, through the relation,
\begin{equation}
L = 4 \pi \, f_{\rm bol} \, {C^2} / {\pi_{\rm p}^2},
% L = 4 \pi f\frac{C^2}{\pi_{\rm p}^2},
\end{equation}
where $C$ is the conversion from parsecs to cm (3.086 10$^{18}$), assuming the bolometric flux is given in cgs units.
The values derived for the effective temperature and luminosity for each case considered are shown in Table~\ref{tab:res}. The corresponding error bars are derived from analytical expressions.

The derivation of a precise value for the effective temperature of 10~Aql was found to be somewhat compromised by the difficulty in establishing a precise value for the bolometric flux of the star. A similar situation was already found for the case of the roAp star $\gamma$~Equ~\citep{2011A&A...526A..89P} and is mainly due to the limited quality of flux-calibrated spectra available in the literature. As an example, in the present case of 10~Aql,  the STIS spectra used in {\it case~1} was acquired with the star offset from slit center by more that 0.9 pixels. According to the notes published in the NGSL library site$^4$, in this case the slit throughput correction applied to the spectra is unreliable. On the other hand, in {\it case~2}, the ELODIE spectrum used in replacement of the STIS spectrum, for wavelengths longer than 3900~\AA, does not have the best quality flag for the flux calibration (ha\-ving a flag of 2 in a scale where the best is 3 and the worst is 0). Given the difficulty in establishing which of the derived bolometric fluxes is the most reliable, we compute evolutionary tracks for both cases.

\begin{figure*}[t]
\begin{center}
\includegraphics[height=67mm]{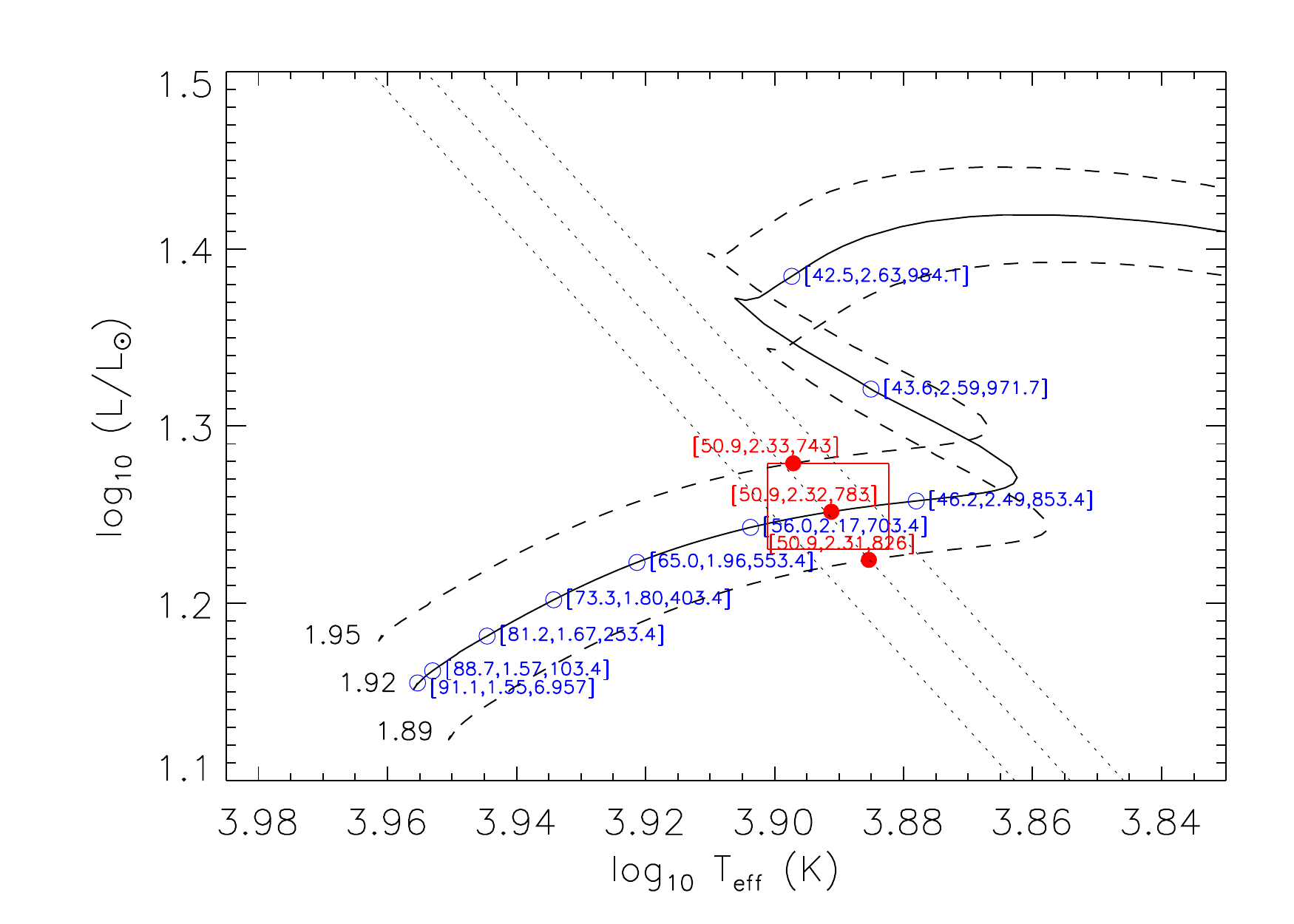}\includegraphics[height=67mm]{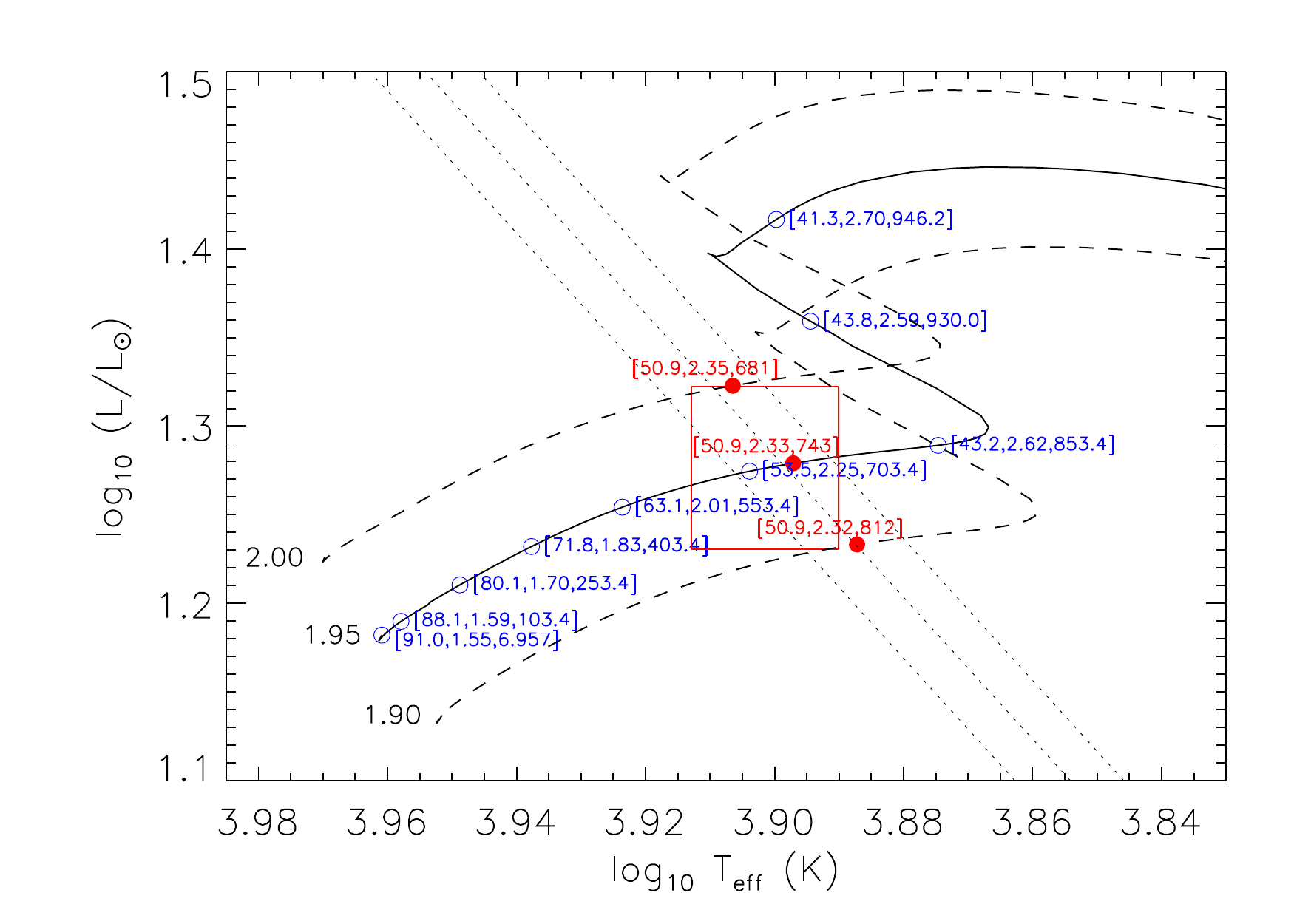}
\end{center}
\caption{The evolutionary tracks corresponding to the {\it case~1} (left) {\it case~2} (right) are represented with the observational error box (the 1$\sigma$-error box ($\log T_{\rm eff}$, $\log L$) in red and the diagonal dotted lines for $R$). The mass (in solar units) is indicated at the beginning of the evolutionary tracks. In each case, the best model is shown as full line, whereas the two extreme models are shown as dashed lines. For the best model, we show in brackets for several time steps (open circles) the evolution of the mean large separation (${\rm \mu Hz}$), the radius (${\rm R_{\odot}}$), and the age (Gy). The full circles represent the time at which the large separation equals the observed one, i.e. $50.9\,\mu Hz$. [See online edition for a color version.]}
\label{hr}
\end{figure*}

\section{Stellar evolution modeling}
\label{CESAM}

We used the stellar evolution code CESAM \citep{morel97,morel08} to derive the mass and the age of the star. The code solves the 1D equations of the  quasi-hydrodrostatic stellar evolution coupled with the detailed OPAL equation of state \citep{roger96} and opacities \citep{roger92}. The nuclear pp- and CNO-reaction rates are determined through the NACRE \citep{Angulo99} tables. The convection is treated using the \citet{canuto91} formalism. The stellar evolution models are coupled with MARCS \citep{gustafsson08} model atmospheres as external boundary conditions. Neither che\-mical element diffusion nor rotation are taken into account in the calculations. Owing to the strong stratification of the chemical elements into the atmospheres of these magnetic stars, the chemical composition is still a matter of debate. However, spectro-photometry observation can be reproduced well with solar metallicity \citep{ryabchikova00}. Moreover, \citet{Huber} have explored several cases of chemical mixture for 10~Aql and found that the solar one reproduces spectro-photometric data well. Therefore, following these authors, we decided to use a solar chemical mixture: Z/X~=~0.024 and Y~=~0.276. Each stellar model is then defined by its mass, age, and convection parameters.

To derive accurate masses and ages, we took advantage of 10~Aql being a multiperiodic pulsator \citep{Huber}, which therefore allows the determination of the large frequency separation $\Delta \nu$ between eigenmodes of consecutive radial orders. This quantity scales as the square root of the mean density of the star $\Delta \nu \propto \sqrt{<\rho>}$  and therefore decreases as the star evolves. Its value has been accurately derived for 10~Aql by \citet{Huber} using the data from the MOST space mission. They find $\Delta \nu$~=~50.95~$\pm$~0.03~$\mu$Hz. The magnetic field creates a shift in the frequencies and therefore affects the large separation. \citet{bigot02} calculated large separations for a field strength of 800~G, close to the upper limit of 1~kG suggested by \citet{ryabchikova00}. Even if such a value of the magnetic field strength leads to a shift in frequency in a few $\mu Hz$, the relative effect between two consecutive modes is weak $\sim 0.1-0.3\, \mu Hz$ and therefore does not influence the derived mass and age, which are mainly constrained by the observational error box in effective temperature, luminosity, and radius.

We considered a null overshoot parameter and various masses between 1.8 and 2.2~$M_{\odot}$. We kept the stellar models whose evolutionary tracks pass through the observational error box $[{\rm T_{eff}, log\,L}]$  and whose mean large separation was equal to the observed one. For our fixed chemical composition, we found two different values of the mass depending on the case considered: $1.92 \pm 0.03$~${\rm M_{\odot}}$ \textit{(case~1)} and $1.95 \pm 0.05$~${\rm M_{\odot}}$ \textit{(case~2)}. These values of masses and ages (given in Table~\ref{tab:res}) and especially their uncertainties must be taken with care considering the standard stellar (nonmagnetic) evolution models used in this paper and the fixed choice of chemical mixture (Z/X, Y).

\section{Discussion and conclusion}
\label{discussion}

By benefiting from the long CHARA baselines and the visible range of the VEGA instrument, we managed to determine an accurate angular diameter for a target smaller than 0.3~mas. Even though 10~Aql is at the angular resolution limit of VEGA, we reach an accuracy of about 3\% on the angular diameter since the recorded fringe contrast is high enough ($V^2 \geq$~0.4)
with hectometric baselines. Using the well-known parallax we derived a linear radius of
$R$~=~2.32~$\pm$~0.09~${\rm R_{\odot}}$. From the spectroscopic data, we computed 10~Aql's bolometric flux and derived its effective temperature (7800-8000~K), luminosity (18 or 19 ${\rm L}_{\odot}$), and mass (1.92-1.95${\rm M}_{\odot}$). All these values can be
compared to the previous determinations found in the literature.

{As regards to the luminosity, the values we derived in this work are in full agreement with those determined by \cite{Nesvacil}. These luminosities are slightly lower than those derived from \cite{Kochukhov}, who found 21.4~${\rm L}_\odot$. This may be due to the bolometric correction adopted by the authors, which was assumed to be the same as for normal stars. Nevertheless, the difference between these values of luminosity is within the error bars.}

{The masses we derived agree with those found by \cite{Huber} (1.95~M$_\odot$), either by calculations from the observed large frequency separation or by model fits based on the pulsation frequencies. But they are larger than those derived in \cite{Nesvacil}, based on their values of $\log g$ and $R$, which span from 1.4 to 1.7~M$_\odot$ according to the spectrophotometric data set they considered.}

From their spectroscopic data and the Warsaw-New Jersey stellar evolution and pulsation code \citep{Pamyatnykh}, and assuming a luminosity of 21.4~${\rm L}_{\odot}$ \citep{Matthews}, \citet{ryabchikova00} derived a mass of 2.0~$\pm$~0.2~$M_{\odot}$ and two values for the stellar radius, depending on the chosen effective temperature: $R$~=~2.50~$\pm$~0.2~R$_{\odot}$ for $T_{\rm eff}$~=~7550~K, and $R$~=~2.38~$\pm$~0.2~R$_{\odot}$ for $T_{\rm eff}$~=~7760~K, respectively. This led them to $\log g$~=~3.95~$\pm$~0.25 and $\log g$~=~3.99~$\pm$~0.25, respectively. They finally adopted $T_{\rm eff}$~=~7550~K and $\log g$~=~3.99~$\pm$~0.25 (and thus $R$~=~2.50~$\pm$~0.2~R$_{\odot}$) since these values were found to reproduce all the observations best. Their values for the radius agree with our linear radius of $R$~=~2.32~$\pm$~0.09~${\rm R_{\odot}}$. Using the same effective temperature ($T_{\rm eff}$~=~7550~K), \cite{Nesvacil} constructed a self-consistent model atmosphere for 10~Aql based on a surface gravity of $\log g$~=~3.8. When fitting their model to spectrophotometric observations, the authors derived a li\-near radius spanning from 2.59~${\rm R}_{\odot}$ to 2.72~${\rm R}_{\odot}$, depending on the assumed atmospheric chemistry (in particular, whether the model is He-solar or He-weak). When their model atmosphere was instead fitted to STIS data ({\it i.e.}, identical to those considered in \textit{(case~2)} of Sect.~\ref{flux}), the authors found, for an effective temperature of $T_{\rm eff}=7550$~K, a slightly smaller radius of 2.46~$\pm$~0.06~R$_\odot$, which is still more than the value derived here. {Concerning the effective temperatures, while our determinations (7800-8000~K) agree with the determination of \cite{Kochukhov} based on photometric data (7925~K), we show a clear discrepancy between our effective temperatures and those adopted either by \citet{ryabchikova00} or \cite{Nesvacil}. Based on spectro-photometry analysis (and especially on the fit of the Paschen continuum), they considered an effective temperature of about 7550~K in their modeling, which is significantly smaller than our determination.}

{Such discrepancies between the \textit{spectroscopic} and the \textit{interferometric} radii and temperatures have been already pointed out in the past for previous studies of roAp stars like $\beta$~CrB, $\alpha$~Cir, and $\gamma$~Equ. In all these cases the radii \textit{(respectively, effective temperature)} derived from optical interferometry were larger \textit{(respectively, lower)} than those derived from spectroscopic methods \citep{shulyak}, which goes the opposite direction from the case of 10~Aql shown in this work. In addition, we carefully checked our systematic errors, such as the calibrator diameter effect, as well as the dependence of our determination on $\log g$, metallicity, and wavelength. As regards wavelength dependency, the angular diameter seemed to be larger (at less than 1-$\sigma$) at shorter wavelengths (around 700~nm), but this trend needs to be confirmed by more accurate data, since we do not have enough visibility measurements to derive accurate angular diameters as a function of wavelength. Our poor (u, v) coverage meant we could not studied the effect of the baseline orientation. It would be very difficult to detect an asymmetry of the angular diameter even with a richer (u, v) coverage because of the very small angular size of the target. As a consequence, it is likely that the discrepancy between our radius and the one derived from atmosphere models is related to model ingredients, so clearly a deeper modeling of 10~Aql is required. Determining accurate and precise fundamental parameters of these complex stars by a method that is as independent as possible of an atmosphere model is thus an efficient way to improve the latter.}

{Finally, we want to point out that the discrepancy is less than 2$\sigma$ on the radius, while there are significant differences found in spectroscopic and photometric determinations of the global parameters of roAp stars, and large differences in effective temperature determined by spectroscopy using models with and without stratification of the elements. For instance, for $\alpha$~Cir, \citet{Kupka} determine 7900~K, while a recent determination by \citet{Koch2009} gives 7500~K. Indeed, models of atmospheres of Ap stars are extremely complex, and the consequent determination of effective temperature (hence radius, assuming one knows the luminosity) by spectroscopic data is subject to major difficulties. A good way to test the different effective temperatures is to use them as input for excitation models since the excitation region is very sensitive to the temperature. Such a study is in progress \citep{cunha13}. A byproduct of these works would be a calibration of the effective temperature scale for these peculiar pulsating stars.}

This work emphasizes the potential of combining interferometric (radius and derived effective temperature) and asteroseismic (large frequency separation) data to improve determination of the mass and the age of stars, as already pointed out by \citet{Creevey}. Our results also clearly demonstrate the feasibility of our roAp program on CHARA/VEGA and the strong interest of increasing the number of targets in our sample. For this program on small targets (usually smaller than 0.5~mas) with the current operational interferometric facilities, only visible long-baseline interferometry can bring accurate angular diameters, and all instrumental improvements towards better sensitivity and better accuracy are of utmost importance.\\

%{\appendix { Appendix: Appendix A}}

\begin{acknowledgements}
The authors are grateful to Denis Shulyak for fruitful discussions
concerning computing the bolometric flux. VEGA is supported
by the French programs for stellar physics and high angular resolution PNPS and
ASHRA, by the Nice Observatory, and the Lagrange Department. The CHARA Array
is operated with support from the National Science
Foundation through grant AST-0908253, the W. M. Keck Foundation, the
NASA Exoplanet Science Institute, and from Georgia State University.
This work has been supported by a grant from LabEx OSUG@2020 (Investissements
d'avenir - ANR10LABX56) and made use of funds from the ERC through the project FP7-SPACE-2012-312844.
MC acknowledges financial support from the FCT through the grant SFRH/BPD/84810/2012.
This research made use of the SearchCal and LITPRO services of the Jean-Marie
Mariotti Center, and of CDS Astronomical Databases SIMBAD and VIZIER.
\end{acknowledgements}

\end{document}